\documentclass[a4,11pt]{article}
\usepackage{amsmath}
\usepackage[round]{natbib}
\usepackage{fullpage}
\usepackage{amsfonts}
\usepackage{graphicx}
\usepackage[usenames,dvipsnames,table]{xcolor}
\usepackage{bm}
\usepackage[colorlinks=true, linkcolor=MidnightBlue, urlcolor=blue, citecolor=cyan]{hyperref}
\usepackage{float}
\usepackage{lineno}
\usepackage{multirow}
\usepackage{url}
\usepackage{authblk}
\usepackage{hyperref}
\modulolinenumbers[5]

\begin{document}
\newcommand{\diff}{\ \mathrm d}
\newcommand{\xpt}{\mathbb E}
\newcommand{\var}[1]{\mathrm{Var}\left[#1\right]}
\newcommand{\parzia}[2]{\frac{\partial #1}{\partial #2}}
\newcommand{\parzib}[2]{\frac{\partial^2 #1}{\partial #2^2 }}
\newcommand{\deriv}[2]{\frac{\diff #1}{\diff #2 }}
\newcommand{\mezera}{\hspace{1.2cm}}
\newcommand{\placeholder}[1]{\newline UOUOUOUOUOUOUOU\newline #1 \newline UOUOUOUOUOUOUOU \newline}
\renewcommand{\appendixname}{Appendix}

\newcommand{\bt}{\boldsymbol\theta}

\def\litID{{\sf id}}
\def\identy{{\mathsurround0pt\mathchoice{\textidenty}{\textidenty}{\scptidenty}{\scptidenty}}}
\def\scptidenty{\setbox0\hbox{$\scriptstyle1$}\bothidenty}
\def\textidenty{\setbox0\hbox{$1$}\bothidenty}
\def\bothidenty{\rlap{\hbox to.97\wd0{\hss\vrule height.06\ht0 width.82\wd0}}
 \copy0\rlap{\kern-.36\wd0\vrule height1.05\ht0 width.05\ht0}\kern.14\wd0}

\noindent\Large{Accelerating Bayesian inference in hydrological modeling with a mechanistic emulator}
\\[2ex]
\normalsize
\noindent\textbf{David Machac}$^{1,2}$, \textbf{Peter Reichert}$^{1,2}$,  \textbf{J{\"{o}}rg Rieckermann}$^{1}$, \textbf{Dario Del Giudice}$^{1,3,4}$and \textbf{Carlo Albert}$^{1}$

\vspace{1cm}



\noindent $^1)$\textit{Eawag, Swiss Federal Institute of Aquatic Science and Technology, Department of Systems Analysis, Integrated Assessment and Modelling, 8600 D\"{u}bendorf, Switzerland}

 \noindent $^2)$\textit{ETH Zurich, Department of Environmental Systems Science, 8092 Zurich, Switzerland}

 \noindent $^3)$\textit{ETH Zurich, Department of Civil, Environmental and Geomatic Engineering, 8092 Zurich, Switzerland}

 \noindent $^4)$\textit{Department of Global Ecology, Carnegie Institution for Science, Stanford, California, USA}

\small

\section*{Abstract}

As in many fields of dynamic modeling, the long runtime of hydrological models hinders Bayesian inference of model parameters from data.
By replacing a model with an approximation of its output as a function of input and/or parameters, emulation allows us to complete this task by trading-off accuracy for speed.
We combine (i) the use of a mechanistic emulator, (ii) low-discrepancy sampling of the parameter space, and (iii) iterative refinement of the design data set, to perform Bayesian inference with a very small design data set constructed with 128 model runs in a parameter space of up to eight dimensions.
In our didactic example we use a model implemented with the hydrological simulator SWMM that allows us to compare our inference results against those derived with the full model. This comparison demonstrates that iterative improvements lead to reasonable results with a very small design data set.
\\[2ex]


Keypoints
\begin{itemize}
  \item Mechanistic emulation
  \item Design data points selection
  \item Calibration of hydrological models
\end{itemize}

\normalsize

\newpage
\section{Introduction}

To summarize, formalize and test our understanding of environmental systems, and to predict the effect of management measures, we need abstract representations of these systems in the form of mental or mathematical models.
In particular, for quantitative predictions, mathematical models are unavoidable tools.
Despite the universality of natural laws, due to the required simplifications of the extreme complexity and diversity of environmental systems, such models contain parameters that need empirical and potentially site- or case-specific calibration.
To adequately address uncertainty in our knowledge, stochasticity in system behavior resulting from intrinsic stochasticity and the effect of unknown and often time-varying influence factors, and remaining systematic deviations of model results from reality, we need care in probabilistically formulating such models.
Calibration then consists of statistical inference of model parameters \citep{kavetski_2006_inputuncertaintyhydrology1,kavetski_2006_inputuncertaintyhydrology2} and, potentially, of stochastic input \citep{del2016inputunc}, intrinsic internal random variables \citep{Reichert2009}, and model bias \citep{kennedy,giudice2013improving,del2015model}.
As there is usually prior knowledge about parameters available from model applications to similar systems, Bayesian inference is the most straightforward methodology to combine this prior knowledge with actual data from a given case study.

Unfortunately, the numerical implementation of Bayesian inference requires many model simulations with different inputs and/or different parameter values.
For models based on large systems of ordinary or partial differential equations, this can be computationally infeasible.

To address this type of problem in computationally demanding tasks, such as inference, sensitivity analysis, or uncertainty propagation, it has been suggested to replace deterministic simulation models by surrogate models that approximate the response surface of the model as a function of input and parameters \citep{kennedy,Bayarri2007a,Conti2009,Castelletti20125}.
When using surrogate models, also called emulators, a reasonable trade-off between accuracy and speed has to be found. This can be achieved by (i) training the emulator on design data points generated by the simulator that are placed strategically in the interesting region of parameter space and (ii) using emulators with a high interpolation accuracy.

Over the past three decades, many data-driven surrogate modeling techniques have been proposed.
Recent overviews, with application to hydrology, are given by \cite{razavi} and by \cite{asher2015review}.
More widely used methods, which found application in hydrology, include artificial neural networks (ANN) \citep{khu2004fast}, polynomial chaos expansion \citep{schobi2015polynomial,laloy2013efficient}, and radial basis functions \citep{bliznyuk2008bayesian}.
A notable disadvantage of these methods is, however, that they do not consider the knowledge about the mechanisms considered by the original model and that they do not provide us with information regarding the uncertainty of their output.

To address theses issues, we focus here on Gaussian Process emulators, as they allow us to get a probability distribution of emulated results to characterize emulation uncertainty \citep{ohag}, and we apply them in a way that allows us to consider our knowledge of the intrinsic mechanisms represented by the original model \citep{Reichert2011,Albert,Machac,emu_appli}.
The uncertainty estimates can be useful when making choices regarding the accuracy-speed trade-off.

It is the goal of this paper to develop an approach to accelerate Bayesian inference for (urban) hydrological models by using an emulator.
The main features of the chosen approach are
(i) to use a likelihood function that accounts (empirically) for systematic errors by applying a statistical bias description technique \citep{kennedy,Bayarri2007a,Reichert2012, giudice2013improving};
(ii) to use a mechanistic emulator to improve interpolation accuracy between design data points \citep{Reichert2011,Albert,Machac,emu_appli};
(iii) combine low-discrepancy sampling \citep{halton_1960_quasirandom,hammersley_1960_montecarlo,hammersley_1964_montecarlo,halton1964algorithm,niederreiter_1992,Reichert2002} with an iterative refinement process for the design data set to get best results with a minimum number of design data points.
As this last point is a new element to similar approaches documented in the literature, it builds the technical focus of this paper.

In the following, we provide more details regarding these three steps and apply the procedure to the simulation of a rainfall-runoff event in an urban catchment with a hydrological model.

\section{Methods}


\subsection{Model likelihood function}\label{sec:likelihood}

Models of complex environmental systems are always biased, due to the inevitable focus on most relevant inputs, simplification of processes, and aggregation of state variables.
This leads to systematic deviations of model outputs from measured data (residuals).
In hydrology, it is still commonplace to ignore such systematic deviations and model residuals with independently and identically distributed errors. This leads to biased parameter estimates and aggravates the bias of the predictions.
It has been demonstrated that adding a simple autoregressive normal bias correction term \citep{kennedy,Bayarri2007a}, in addition to i.i.d. errors, to the output of a hydrological model can greatly reduce the bias in parameter estimates and lead to more reliable predictions \citep{Reichert2012, giudice2013improving}.

Here, we consider dynamical models, whose outputs are univariate time-series described by random vectors of the form

\begin{align}\label{model}
	\mathbf Y(\boldsymbol\theta,\sigma_E,\sigma_B,\tau)
=
g^{-1}\Bigl(
    g\bigl({\mathbf y}(\boldsymbol\theta)\bigr) + \mathbf B(\sigma_B,\tau)+ \mathbf E(\sigma_E)
\Bigr)
\,.
\end{align}
The deterministic model output ${\mathbf y}(\boldsymbol\theta)$ is a time-series that depends on model parameters $\boldsymbol\theta$. Its components, $y_i(\boldsymbol\theta)$, for $i=1,\dots,N_t$, are associated with measurements at time points $t_i$.
The third term on the r.h.s., $\mathbf E(\sigma_E)$, denotes the measurement error white noise with zero prior mean and standard deviation $\sigma^2_E$, and the second term $\mathbf B(\sigma_B,\tau)$ is the additive bias correction term with zero prior mean and covariance matrix given by

\begin{align}
\Sigma_{B,i,j}(\sigma_B,\tau)=\sigma^2_B\exp\left(-\frac 1{\tau}\left|t_i-t_j\right|\right)\,.
	\label{eqn:bias}
\end{align}
This bias correction term is parameterized by its standard deviation $\sigma_B$ and auto-correlation time $\tau$.
The function $g$ (applied point-wise to the time-series) is a transformation that is used to account for the ubiquitous hetero-scedasticity in hydrological data. A typical choice is the Box-Cox transformation \citep{box1964analysis}, $g(y)=(y^\lambda-1)/\lambda$, for $\lambda\neq 0$.

The logarithm of the associated likelihood function, evaluated at a measured time-series, ${\mathbf y_o}$, reads

\begin{multline}
	\ln l(\mathbf y_o|\boldsymbol\theta,\sigma_E,\sigma_B,\tau)
	     =  - \frac{N_t}{2}\ln \left|\sigma_E^2\boldsymbol 1+\Sigma_B\right|
 \\
 - \frac{1}{2}\Bigl(g(\mathbf y_o)-g\bigl({\mathbf y}(\boldsymbol\theta)\bigr)\Bigr)^T\bigl(\sigma_E^2\boldsymbol 1+\Sigma_B\bigr)^{-1}\Bigl(g(\mathbf y_o)-g\bigl({\mathbf y}(\boldsymbol\theta)\bigr)\Bigr)
+\operatorname{const}
\,.
	     \label{eqn:lhood}
\end{multline}
where $\mid .. \mid$ represents the determinant of the enclosed matrix.
For more details on the motivation behind this likelihood and its derivation, the reader is referred to the cited literature.

\subsection{Model calibration through Bayesian inference}

Bayesian statistics provides a mathematical framework for updating prior knowledge or belief about model parameters with information from data through a consistent learning process.
If model parameters have a physical meaning, such as in the model used in Sect. 3, we typically have some prior knowledge about their values, which we encode in terms of prior probability distributions. Furthermore, the prior for the standard deviation of the bias correction term $\sigma_B$ is used to express our desire that the data is predominantly explained by the model and not by the correction term.
As $\sigma_E$ is predominantly determined by measurement noise, we derive a prior from our knowledge of the measurement process.
For the auto-correlation time $\tau$ of the bias correction term we typically use a sharp prior reflecting the characteristic memory time of the model, as this is usually well known and difficult to infer from data.
Combining these marginal priors usually under the assumption of independence, we get the complete description of our prior knowledge as a joint probability distribution of all parameters $f_{\text{prior}}(\boldsymbol \theta, \sigma_E,\sigma_B,\tau)$.
If measured data $\mathbf y_o$, which is believed to be a realization of model (\ref{model}), is available, the posterior, expressing the combined knowledge from prior information and data, is expressed through the Bayesian update rule

\begin{align}
	f_{\text{post}}(\boldsymbol \theta,\sigma_E,\sigma_B,\tau|\mathbf y_o)
    \propto 
    f_{\text{prior}}(\boldsymbol \theta,\sigma_E,\sigma_B,\tau)
    l(\mathbf y_o|\boldsymbol \theta,\sigma_E,\sigma_B,\tau) \,,
	\label{eqn:bayes}
\end{align}
where the unknown proportionality constant is defined by normalization.
Standard techniques used to draw a parameter sample from the posterior, which can then be used, e.g., for making probabilistic predictions, are variants of the Metropolis or Metropolis-Hastings Markov Chain Monte Carlo (MCMC) algorithms.
Here, we are using the \textsc{emcee} Python package \cite{foreman2013emcee}, which implements an ensemble method that runs several interacting Markov chains in parallel \citep{goodman2010ensemble}.
Despite the relatively low autocorrelation of this algorithm, a few thousand or tens of thousands of evaluations of the likelihood function, for different parameter sets, are required. As each evaluation of the likelihood function requires a model run, many simulators used in the environmental sciences are simply too slow to allow for a full-fledged Bayesian inference with these techniques.

Therefore, we suggest to replace the simulator by a much faster, yet less accurate emulator, for the Bayesian inference procedure as introduced by \cite{kennedy}.
The aim of this paper is to design an inference algorithm that manages with as few simulator runs as possible.
To this end, we propose to use a mechanistic emulator conditioned on a Halton sequence of design parameter vectors, as outlined in the next subsection.
An iterative improvement of this emulator within the Bayesian inference procedure is outlined in Sect. \ref{sec:locref}.

\subsection{Mechanistic, dynamic emulators}\label{sec:emulator}

\begin{figure}
 	\centering
 	\includegraphics[scale=0.6]{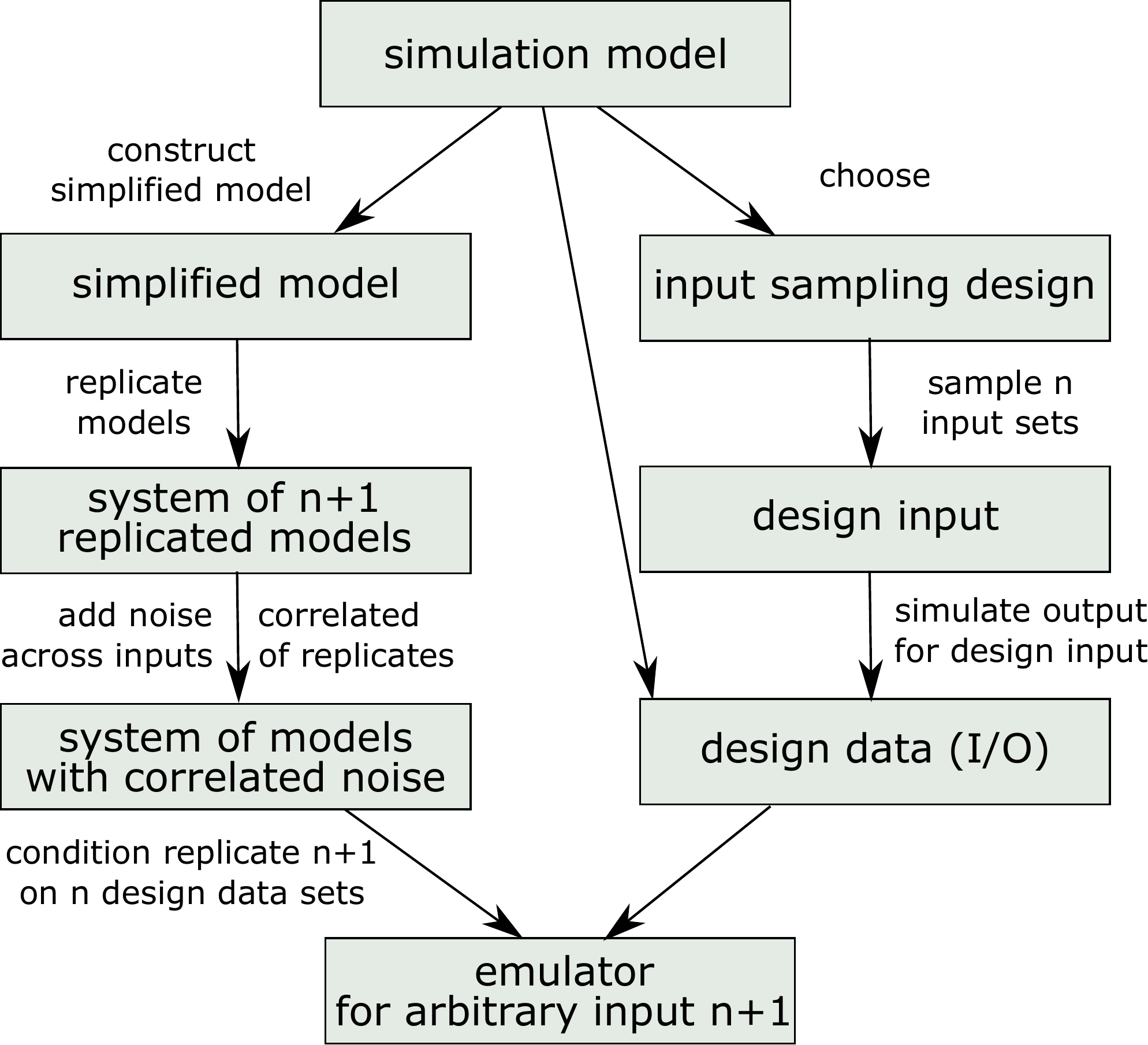}
     \caption{Schematic visualization of the construction process of the emulator. The left part illustrates the development of the equations and has only to be done once. The right part represents the steps needed for conditioning the emulator to simulated design data. This part has to be redone, whenever more simulation data become available.}
 	\label{fig:emulator_construction}
 \end{figure}

The emulator we are using in this work is a particular kind of stochastic approximation to a simulator, interpolating the response surface of the simulator between design input-output pairs - the design data that was generated with the simulator. In our case, we are interested in dynamic emulators that take parameter vectors as inputs and generate time-series as outputs.
Our emulator is based on a Gaussian prior that is conditioned on design data. Compared to standard emulators, {\em mechanistic} emulators use our knowledge of simulator processes to define better priors that require less design data points to achieve a satisfactory approximation of the simulator \citep{Reichert2011}.
The construction of a mechanistic, dynamic emulator proceeds in the following 5 steps (see also Fig.\ \ref{fig:emulator_construction}:
\begin{enumerate}
  \item
  Find an adequate low-dimensional state space that simplifies the state space of the simulation model, and an adequate system of linear ordinary differential equations (ODE) on it that approximates, to lowest order, the dynamics of the simulation model. This linear ODE may depend non-linearly on parameters of the simulation model as well as on additional auxiliary parameters.
  \item
  Add normal white noise to this linear model compensating for all the omissions in the simplification process.
  For conditioning of this noise to the design data, couple $n+1$ replica of the resulting stochastic linear model, for $n+1$ different parameter sets, through the noise term in such a way that the closer the parameter vectors are the stronger is the coupling.
  \item
  Generate $n$ input-output pairs (design data) with the simulator.
  \item
  Determine the auxiliary emulator parameters through maximizing the normal likelihood function of the first $n$ replica, evaluated at the design data points.
  \item
  Condition the coupled system of $n+1$ replica to the design data. The result is a normal distribution that encodes the best guess and uncertainty estimation, for the output time-series of the simulator, evaluated at the $(n+1)^{\mathrm{th}}$ parameter set.
\end{enumerate}

To avoid confusion it is very important to distinguish the two different models and likelihood functions that play a role here. On the one hand, we have the {\em simulation model} (\ref{model}), with associated likelihood function (\ref{eqn:lhood}). Together with {\em measured data} $\bf y_0$ it is used to update our prior knowledge about model and noise parameters $f_{prior}(\boldsymbol \theta, \sigma_E,\sigma_B,\tau)$. On the other hand, we use a {\em simplified prior model} as described in points (1) and (2) above. Its deterministic part is a very crude linear approximation of the dynamic simulation model that is used to calculate $\bf y(\boldsymbol \theta)$. It can depend in a non-linear fashion on the model parameters $\boldsymbol\theta$ and on auxiliary parameters. The likelihood function of the (coupled) simplified prior model is normal (because it is derived from a linear state-space model with additive normal noise) and has nothing to do with the likelihood function of the simulation model. It is used to estimate the auxiliary parameters of the prior model using {\em simulated design data}. After estimating the auxiliary parameters and conditioning to design data, the prior model turns into an {\em emulator} - a fast statistical approximation of the slow simulator, which can then be used instead of the emulator, for simulation-intense tasks such as Bayesian inference.

Depending on the simulation model, the mechanistic emulator could be replaced by a purely data-driven surrogate with similar or even better performance \citep{Carbajal_2017}. The iterative improvement for the purpose of Bayesian parameter inference would work just as well with such surrogates. 
In this paper, we restrict ourselves to mechanistic emulators. In the remainder of this section, we briefly explain the mathematics behind points 1-5 described above. Readers who are not familiar with emulators are referred to \cite{Albert} or \cite{Machac}, for more in-depth explanations.  

\subsubsection{Simplified linear model}

Urban hydrological systems are typically modeled by means of systems of ODEs on high-dimensional state spaces of interconnected linear and non-linear reservoirs.
For systems with univariate output time series, previous studies have shown that using a single linear reservoir as a prior for the emulator already leads to drastically increased accuracy compared to standard non-mechanistic emulators \citep{Machac,emu_appli}.
Thus, as a prior for our emulator, we use a single state variable, $d(t)$, measuring water height in a linear reservoir, and model its dynamics by means of eq.

\begin{align}
    \frac {\ \mathrm d d(t)} {\ \mathrm d t} = \kappa\left(\boldsymbol\theta,\boldsymbol\theta'\right)d(t)+p(t,\boldsymbol\theta,\boldsymbol\theta'),
    \label{eqn:simpMod}
\end{align}
where $\kappa\left(\boldsymbol\theta,\boldsymbol\theta'\right)$ is a function of the simulator parameters $\boldsymbol\theta$, and of as yet unspecified auxiliary emulator parameters $\boldsymbol\theta'$, and $p(t,\boldsymbol\theta,\boldsymbol\theta')$ is associated with the rain input into the system.
The output time-series of the linear model is derived from the state variable through the linear relationship

\begin{align}
  y_i=h(\boldsymbol\theta,\boldsymbol\theta')d(t_i)\,,
	\label{eqn:projection_}
\end{align}
where $h(\boldsymbol\theta,\boldsymbol\theta')$ is a function of model parameters and may depend on auxiliary emulator parameters as well. There is no recipe how to design the functions $\kappa$, $p$ and $h$. Anything that leads to a mapping of model parameters $\boldsymbol\theta$ to model outputs $\bf y$ resembling the simulation model $\bf y(\boldsymbol\theta)$ is allowed. The better the resemblance the less design data will be needed for an accurate emulation. An example will be given in Sect. \ref{sec:caseStudy}.

\subsubsection{Couple $n+1$ replica through a noise term}

To make up for the simplification inherent to the linear model introduced in the previous paragraph, we extend it with a Gaussian process conditioned on design data.
To this end, we couple $n+1$ replica of the linear model \eqref{eqn:simpMod} by means of a multivariate normal noise term according to eq.

\begin{align}
	\frac {\ \mathrm d \mathbf d(t)} {\ \mathrm d t} = \boldsymbol{\kappa}\left(\widetilde{\boldsymbol{\theta}},\boldsymbol{\theta}'\right)\mathbf{d}(t)+\mathbf p(t,\widetilde{\boldsymbol{\theta}},\boldsymbol{\theta}')+ \mathbf C\left(\widetilde{\boldsymbol \theta},\boldsymbol{\theta}'\right)\boldsymbol{\eta}(t),\;\; {\mathbf d}\in\mathbb R^{n+1},
	\label{eqn:coupled_sys}
\end{align}
which is a system of stochastic linear differential equations (SDE).
We distinguish different replica by Greek indices and define tensors $\boldsymbol \kappa$ and $\mathbf p$ by eqs.

\begin{align}
	\boldsymbol\kappa^{\alpha}_{\beta}\left(\widetilde{\boldsymbol \theta},\boldsymbol{\theta}'\right)
=
\delta^{\alpha}_{\beta}
\kappa\left(\boldsymbol\theta^{\alpha},\boldsymbol\theta'\right) \;\; \text{and}\;\;\mathbf p^{\alpha}(t,\widetilde{\boldsymbol\theta},\boldsymbol\theta')=p(t,\boldsymbol\theta^{\alpha},\boldsymbol\theta'),
\end{align}
where $\delta_\beta^\alpha$ is the Kronecker delta function or the identity matrix.
Vector $\mathbf d:=\left(d_1,\dots,d_{n+1}\right)$ is the state of the coupled system and $\widetilde{\boldsymbol\theta}:=\left(\boldsymbol\theta^1,\dots,\boldsymbol\theta^{n+1}\right)$ denotes the different parameter vectors associated with the $n+1$ replica.
The noise term $\mathbf C(\widetilde{\boldsymbol \theta},\boldsymbol{\theta}')\boldsymbol{\eta}(t)$ in equation \eqref{eqn:coupled_sys}, where $\boldsymbol\eta$ denotes Gaussian white noise, couples the replica. The coupling matrix is defined by eq.

\begin{align}
	\mathbf C\left(\widetilde{\boldsymbol\theta},\boldsymbol{\theta}'\right) = \sigma\mathbf R\left(\widetilde{\boldsymbol\theta},\boldsymbol{\theta}'\right),
        \label{eqn:noiseterm}
\end{align}
where $\sigma$ is the standard deviation of the noise and part of the auxiliary parameter vector $\boldsymbol\theta'$. For the square of the correlation function $\mathbf R$, we choose

\begin{align}
        \left(\mathbf R\mathbf R^T\right)^{\alpha\beta}(\widetilde{\boldsymbol\theta},\boldsymbol{\theta}')=\exp{\left(-\frac{1}{\gamma}\sqrt{\sum_{l}\left(\frac{\theta_l^\alpha-\theta_{l}^{\beta}}{\rho_l}\right)^{2}}\right)}\,,
\label{eqn:cor_3}
\end{align}
so that the closer to each other the parameter vectors ${\boldsymbol\theta}^\alpha$ and ${\boldsymbol\theta}^\beta$ are, the stronger the coupling between the associated replica is.
We use a normalizing constant $\rho_l$, which is the span of the calibration hypercube in dimension $l$.
The \textit{correlation length} $\gamma$ is added to the replica non-specific parameters $\boldsymbol\theta'$ and has to be estimated.

For the sake of computational efficiency, it might be beneficial to replace \eqref{eqn:cor_3} by a correlation function with compact support \citep{kaufman2011}. Since we use relatively small design data sets, we do not pursue this approach here.

All the output time series (\ref{eqn:projection_}), for all the $n+1$ replica, which are generated by the system of coupled linear SDEs (\ref{eqn:coupled_sys}) are distributed according to a $(n+1)N_t$ dimensional normal distribution. That is, there is correlation both across time points and across replica. As explained in detail in \citep{Machac}, mean and covariance matrix of this distribution are expressed in terms of the  Green's functions, $ G^\alpha(t',t,\boldsymbol{\theta}')$, of the operators $d/dt - {\boldsymbol\kappa}(\boldsymbol\theta^\alpha,\boldsymbol\theta')$, for all $n+1$ replica, and read, respectively, as

\begin{align}
         z^\alpha(t_i,\boldsymbol{\theta}')
   &=
   h({\boldsymbol\theta}^\alpha,{\boldsymbol\theta}')
   \int G^\alpha(t_i,t,\boldsymbol{\theta}')p(t,\boldsymbol\theta^\alpha,\boldsymbol\theta')dt
   \,,
   \label{eqn:meancoupled}\\
   \Sigma^{\alpha\beta}(t_i,t_j,\boldsymbol{\theta}')
   &=
   \sigma^2
    h({\boldsymbol\theta}^\alpha,{\boldsymbol\theta}')h({\boldsymbol\theta}^\beta,{\boldsymbol\theta}')
    ({\bf R}{\bf R}^T)^{\alpha\beta}(\widetilde{\boldsymbol\theta},\boldsymbol{\theta}')
    \int
    G^\alpha(t_i,t,\boldsymbol{\theta}')
    G^{\beta\dag}(t,t_j,\boldsymbol{\theta}')
    dt
   \,.
   \label{eqn:Sigmacoupled}
 \end{align}

\subsubsection{Design data generation}
We generate parameter vectors $\{\boldsymbol\theta^\alpha\}_{\alpha=1}^n$, for the design data set, with a \textit{Halton sequence} \citep{halton1964algorithm}, which is a \emph{low-discrepancy} sequence \citep{press1996numerical} with the advantage that its points are close to being \textit{equidistributed} (they cover the parameter space evenly).
Yet this sampling method avoids the unwanted ``collapsing'' property of regular grids \citep{urban2010comparison} (which are equidistributed) or the clustering issues with small random samples \citep{santner2013design}.

The parameter set should cover the whole region of interest of the parameter space, which is in our case determined by the calibration bounds. If possible, it should outreach this space of interest, to ensure good accuracy of the emulator on its borders.

The design data is the set of pairs $\{(\boldsymbol\theta^\alpha,{\bf y}^\alpha)\}_{\alpha=1}^n$, where ${\bf y}^\alpha$ is the output time-series of the simulator, for the parameter vector $\boldsymbol\theta^\alpha$.

\subsubsection{Estimation of auxiliary emulator parameters}
\label{subs:aux}

The auxiliary emulator parameters $\boldsymbol\theta'$ can be estimated by maximizing the normal likelihood function of the first $n$ replica, evaluated at the design data points.
In practice, however, it turns out to be difficult to estimate the auxiliary parameter $\gamma$ characterizing the correlation length.
Therefore, we tune this parameter manually, as shown later in the text.
For the estimation of the other auxiliary parameters of the linear model we ignore correlations both in time and in parameter space, and simply minimize the sum of squares between solutions of the linear model and the design data. This is numerically much faster than maximizing the normal likelihood of the prior model and should lead to similar results, considering that in our application the design data points are equally spaced both in time and in parameter space (initially; auxiliary parameters are not re-calibrated during the iterative procedure described in the next chapter).
The auxiliary noise parameter $\sigma$ is important for the estimation of the emulator accuracy only. 
Once all the other auxiliary emulator parameters are estimated, $\sigma^2$ is calculated as follows \citep{Machac}
\begin{equation}\label{eqn:sigmasq_estimate}
\sigma^2=\frac{1}{nN_t}(\bf y-\bf z)^T\Sigma^{\star -1}(\bf y-\bf z)\,,
\end{equation}
where $\bf z$ is defined in (\ref{eqn:meancoupled}) and $\Sigma^\star$ is derived from (\ref{eqn:Sigmacoupled}), by considering the first $n$ replica only, and stripping off the pre-factor $\sigma^2$. 

\subsubsection{Conditioning of the emulator}

The output of our emulator is a multivariate normal distribution estimating the output time-series of the simulator, for the $(n+1)^{\mathrm{th}}$ parameter vector. It is derived through conditioning of the multivariate normal distribution of the coupled system to $n$ design data points.
Mean and covariance matrix of the resulting $N_t$ dimensional normal distribution are calculated as

\begin{align}
        \bar {\mathbf y} &= \mathbf z^{n+1}+\Sigma^{n+1,\alpha}(\Sigma')^{-1}_{\alpha\beta}(\mathbf y^{\beta}-\mathbf z^{\beta}),\label{conditionedmean}\\
        \overline\Sigma &= \Sigma^{n+1,n+1}-\Sigma^{n+1,\alpha}(\Sigma')^{-1}_{\alpha\beta}\Sigma^{\beta,n+1}\label{conditionedcov},
\end{align}
where $\mathbf z$ and $\Sigma$ denote mean and covariance matrix of the unconditioned system, eqs. (\ref{eqn:meancoupled}) and (\ref{eqn:Sigmacoupled}), respectively, and $\Sigma'$ is the covariance matrix associated only with the first $n$ replica. For more details on this procedure, see \cite{Albert} or \cite{Machac}.

\subsection{Iterative, local refinement of the design data set}\label{sec:locref}

\begin{figure}
 	\centering
 	\includegraphics[scale=0.5]{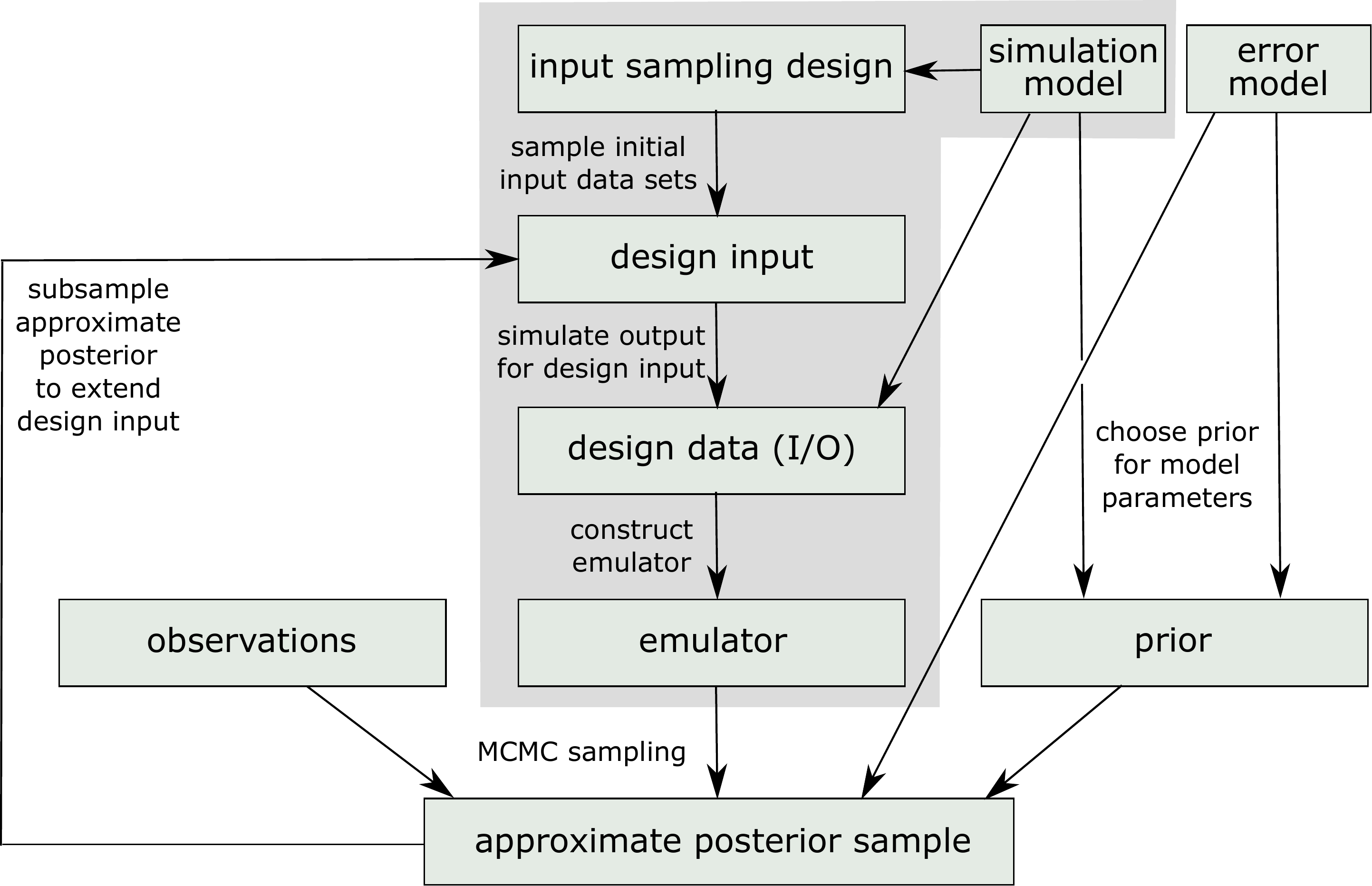}
     \caption{Schematic visualization of the refinement process of the emulator in the context of Bayesian inference. The boxes on a gray background represent the right part of Fig.\ \ref{fig:emulator_construction}. The remainder of the diagram demonstrates essential elements and steps for emulator refinement and Bayesian inference. See text for more details.}
 	\label{fig:emulator_refinement}
 \end{figure}

Initially, the parameter region within which the emulator has to be conditioned can be chosen as a hypercube defined by parameter intervals adjusted to the highest probability density regions of the prior marginals.
This region should then be covered by design parameter vectors as evenly as possible to ensure an efficient use of information from the design data (unless we have specific knowledge of strongly nonlinear behavior of results as a function of parameters in certain parameter regions that should then be covered more densely than other regions).
To do this, we chose to use the low-discrepancy Halton sequence for this initial parameter sample.
However, if the highest probability density region of the posterior is much smaller (because there is a considerable gain of information from the data), these parameter sets are not very efficient to ensure a good accuracy of the emulator in the region relevant for the posterior.
As we do not know a priori where the highest probability density region of the posterior is located, we start with design data covering the prior and iteratively add additional parameter sets by using the information we gained about the posterior from the emulator conditioned to the previous design data set.
This leads to the following empirical procedure to choose $n$ parameter vectors to construct design data for Bayesian inference (see also Fig. \ref{fig:emulator_refinement}):
\begin{enumerate}
	\item Construct $n/2$ design parameter vectors with the Halton sequence from a hypercube defined by initial parameter intervals that are not much larger than the intervals within which there is considerable marginal prior probability. Construct the corresponding design data set by running the simulator for these parameter vectors. Sample from the approximate posterior distribution of the parameters by running the MCMC scheme with the emulator constructed by conditioning its prior to these design data.
	\item Sample $n/8$ data points from this approximate posterior, stretch the sample from its center of mass to cover a somewhat larger parameter region (to account for the approximate nature of the posterior), and run the simulator to get the corresponding model results.
    \item Add these $n/8$ design data points to the original set and condition the emulator to the extended design data.
        The auxiliary emulator parameters $\boldsymbol\theta'$ are not re-calibrated.
	\item Sample from the better approximation to the posterior by running the MCMC scheme with the improved emulator.
	\item Repeat steps 2 to 4 four times to reach the final size of the design data set of $n$.
        \item Assess convergence by checking whether the difference between successive posterior approximations decreases. This can be assessed by applying the distribution-free test for comparing samples from two multivariate distributions by \cite{Rosenbaum_2005}.
        \item Stop if the sequence of the posterior approximations with design data sets of size $n/2$, $5n/8$, $3n/4$, $7n/8$, and $n$ show adequate convergence. Otherwise proceed with larger design data sets or explore different schemes.
\end{enumerate}
Stretching of the sample in step 2 is done by replacing the set of parameter vectors $\{\boldsymbol\theta_i\}$ by

\begin{align}
	\boldsymbol\theta^*_i = \boldsymbol \mu + \lambda(\boldsymbol\theta_i-\boldsymbol \mu) \quad \forall i\quad,
\label{eqn:stretch}
\end{align}
where $\mu$ is the mean of $\{\boldsymbol\theta_i\}$ and $\lambda$ is the \emph{stretch factor}, which we set to $\lambda=1.1$, in order to sample design parameters slightly beyond the approximate posterior. The procedure outlined above ensures an iterative improvement of the emulation accuracy in the parameter region relevant for the posterior.

\subsection{Quantifying differences across samples} \label{sec:crossmatch}

To better quantitatively assess the differences between successive iterative approximations to the posterior as well as between the iterative approximations and the posterior calculated without using the emulator, we applied the distribution-free technique by \cite{Rosenbaum_2005}.
The concept of this approach is to calculate the nearest distances between pairs of points within and across two random samples and to calculate the number of ``cross-matches'', $n_{\mathrm{cm}}$, for which one point of the pair belongs to one sample and the other point belongs to the other sample.
In case of the same sample size, $n$, we would expect that about half of the pairs would be cross-matches: $n_{\mathrm{cm}}=n/2$ and that this number would decline with increasing difference between the distributions.
Thus, we can define a cross-match-distance as

\begin{equation}
   d_{\mathrm{cm}} = 1 - \frac{2n_{\mathrm{cm}}}{ n} \quad ,
\label{equ:crossmatchdistance}
\end{equation}
the expected value of which would be zero for samples from identical distributions and which would reach unity, for completely separated distributions.

\section{Case study}
\label{sec:caseStudy}

In this section we apply the algorithm outlined in the previous section, for the calibration of a Storm Water Management Model (SWMM), which was set up for a medium-sized urban catchment in Switzerland. For didactical reasons we calibrate the model to a single rain event only. This keeps the runtime of the full model at reasonable 19 seconds, which allows us to do the calibration with the full model as well, for comparison with the emulation-based inference. Calibration with the full model takes approximately 6 days on a 24-core, Intel(R) Xeon(R) CPU L5640  @ 2.27GHz system.

\subsection{Catchment and calibration data}

\begin{figure}
 	\centering
 	\includegraphics[scale=0.5]{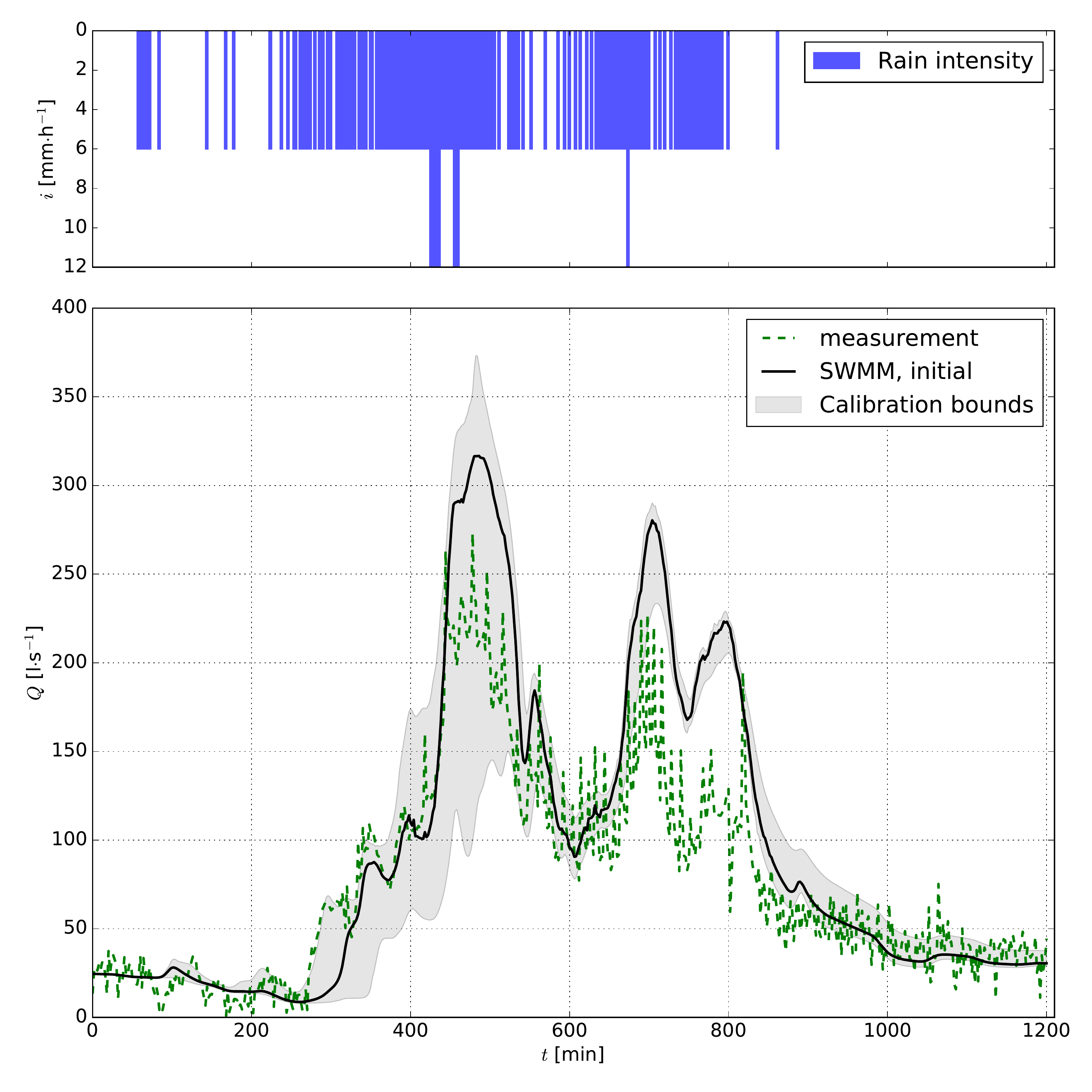}
     \caption{Uncalibrated model and data, for the rain event we use for calibration. The solid line is the SWMM output at the maximum of the prior. The filled area represents the prior parametric uncertainty. One sees that the measurements lie within this area in the first half of the observed time span, but not in the second. This discrepancy will be described by a stochastic bias process.}
 	\label{fig:design_}
 \end{figure}

The catchment is located in the city of Adliswil in the canton of Zurich, Switzerland.
It spreads on both banks of the river Sihl, but we will consider only the part on the right bank.  The area of the catchment is 162.8 ha and it is mainly urban, with about $1/3$ consisting of parks and similar pervious areas.
For calibration of the SWMM model introduced in the next section, we use a single discharge time-series measured at the outlet to the wastewater treatment plant (WWTP).
The time-series has a temporal resolution of two minutes and is derived from contact-free Flo-Dar measurements \citep{flodar}.
The discharge was measured during a single rain event, which occurred on May 28, 2013 and lasted for approximately 15 hours.
Using a single rain event for SWMM calibration is not unusual \citep{knighton2014development}. For us, the main reason behind using a single event was to be able to calculate the exact parameter inference result with the full SWMM model and assess the accuracy of the approximate procedure.
The precipitation, measured by a pluviometer based on the weighing principle (manufacturer: OTT Hydromet GmbH), was very mild, but steady.
We emphasize, that we use a single pluviometer for the whole catchment, which is likely to add to the bias of the model results and justifies the use of a bias correction term in the model.
Precipitation measurements and the associated outflow are shown in Figure \ref{fig:design_}.

\subsection{SWMM model and its parameter vector}
\label{sec:swmmmodel}

The SWMM model for the Adliswil catchment was constructed based on GIS information, as described in great detail in \cite{rao}.
The model is divided into 101 \emph{subcatchments} interconnected through 456 \emph{pipes}. Each subcatchment and each pipe comes with its own parameter vector.
To reduce the number of parameters that need to be calibrated, it is a standard procedure to combine similar parameters, like all of Manning's roughness coefficients of all the pipes, into parameter classes and calibrate a single scaling factor per class only.
The eight classes that we use, along with feasible ranges for the scaling factors, are described in Table \ref{tab:swmmparameters}.
\begin{table}
\begin{center}
\begin{tabular}{l|l|l}
Parameter & Range & Description \\
\hline
$\theta_{Impervious\;area}$ &
[0.5,1.1] &
\parbox[t]{10.0cm}{Scaling factor for the \emph{percentages of impervious area} $[\%]$ for all subcatchments. The average of the initial values, weighted with the areas of the subcatchments, amounts to $36\%$.
The upper bound ensures that none of the individual parameters exceeds $100\%$.} \\
\hline
$\theta_{Width}$ &
[0.5,1.5] &
\parbox[t]{10.0cm}{Scaling factor of the \emph{characteristic widths of the overland flow paths} [m] that determine the response times of the subcatchments. The wider this characteristic width, the faster the response. The initial weighted average is $35.7$ m.
        We refer to the user manual \citep{Huber1988}, for the mapping of subcatchments to the rectangles used in SWMM.} \\
\hline
$\theta_{Slope}$ &
[0.5,1.5] &
\parbox[t]{10.0cm}{Scaling factor of the \emph{slopes of the subcatchments $[\%]$}. The slopes are aggregated measures derived from the elevation map. The initial weighted average is $11.4 \%$.} \\
\hline
$\theta_{Stor.\;imp.}$ &
[0.5,1.5] &
\parbox[t]{10.0cm}{Scaling factor of the \emph{heights of the depression storages on impervious areas} [mm]. The heights are set to $2$ mm in all subcatchments.} \\
\hline
$\theta_{n_{imp}}$ &
[0.5,1.5] &
\parbox[t]{10.0cm}{Scaling factor of the \emph{Manning coefficients for the impervious areas} [s$\cdot$m$^{-1/3}$] that determine the roughness of the impervious parts of the subcatchments.
The initial values are all set to $0.12$ s$\cdot$m$^{-1/3}$.} \\
\hline
$\theta_{Stor.\;per.}$ &
[0.5,1.5] &
\parbox[t]{10.0cm}{Scaling factor of the \emph{heights of the depression storages on the pervious areas} [mm]. These heights are also set to $2$ mm for all subcatchments.} \\
\hline
$\theta_{Imp.\;area\;w/o.\;dep.\;stor.}$ &
[1.0,1.5] &
\parbox[t]{10.0cm}{Scaling factor of the \emph{percentages of impervious area without depression storages} [\%]. The percentages of impervious are have a weighted average of $19.04 \%$. Unlike for other parameters, we use a lower boundary equal to one for the scaling factor due to numerical instabilities arising for smaller values.} \\
\hline
$\theta_{n_{con}}$ &
[0.5,1.5] &
\parbox[t]{10.0cm}{Scaling factor for the \emph{Manning coefficients for the sewer pipes} [s$\cdot$m$^{-1/3}$]. For all pipes, the initial value is set at 0.012 s$\cdot$m$^{-1/3}$.} \\
\end{tabular}
\end{center}
\caption{Scaling factors used as model parameters for calibrating the SWMM model.}
\label{tab:swmmparameters}
\end{table}


We calibrate the model for 2, 4 and 8 parameters, using parameters 1.--2., 1.--4.\ and 1.--8., respectively.
The error model we use, for the Bayesian inference, is described in Sect. \ref{sec:likelihood}. For $g$ we use the \emph{Box-Cox} \citep{box1964analysis} transformation to stabilize the variance. We set $\lambda=0.35$, as suggested in \cite{giudice2013improving}.

\subsection{Error model}
\label{sec:errormodel}

The SWMM model outlined in the previous section is deterministic; it does not consider the structural or output errors and often input uncertainty is not propagated through deterministic hydrological models.
Real systems, at the aggregation level where we can describe them, are not entirely deterministic.
For hydrological systems, the main reasons for this are the use of a spatially and temporarily aggregated state description and the use of aggregated and extrapolated input.
As we cannot describe the system in all details, the same observed and described state and input of the system represents multiple possible underlying true states and multiple amounts and temporal and spatial distributions of input.
This leads to different time evolution of the underlying system -- i.e.\ to non-deterministic behavior.
Ideally, this would be accounted for by considering input uncertainty explicitly \citep{del2016inputunc} and by making time evolution of the model stochastic, e.g.\ by conserving mass balances in making processes, rather than states, stochastic \citep{Reichert2009}.
As errors in input and processes are propagated through the states of the system to the output, such deficiencies in description lead to correlated errors in model output.

In many applications of environmental modeling, a detailed description and propagation of these errors is computationally too demanding.
To still consider the effect of these uncertainties on model results, Kennedy and O'Hagan suggested to introduce a stochastic, autocorrelated bias correction term to the output and to infer its time course jointly with the model parameters \citep{kennedy}.
When using an additive bias correction term and applying a transformation to account for the heterodasticity of the uncertainty \citep{Reichert2012, giudice2013improving}, this leads to our model equation \eqref{model}
with a parameterization of the correlation structure of the bias correction term $\mathbf{B}$ given by equation \eqref{eqn:bias}.
Assuming normal distributions for the bias and observation errors then leads to the likelihood formulated by equation \eqref{eqn:lhood}.

The use of this likelihood has the following advantages over neglecting model bias:
(i) the consideration of the bias correction term leads to better uncertainty estimates of the model parameters and to the identification of the statistical properties of the bias and its time course during the calibration phase.
(ii) the statistical properties of the bias and the identification of its state allow us to consider the uncertainty in the bias for calculating predictive uncertainty bounds of future predictions (that state is only relevant for short-term predictions in the order of the time scale of the correlation time of the bias, after that period, only the statistical properties of the bias are needed).

More technical details and motivation for this kind of bias description is provided in \cite{kennedy} and \cite{Reichert2012},
examples on hydrological applications can be found in \cite{giudice2013improving} and \cite{del2015model}.

\subsection{Selecting a suitable prior}

The prior probability distribution $f_{prior}(\boldsymbol\theta,\sigma_E,\sigma_B,\tau)$ appearing in eq. (\ref{eqn:bayes}), is defined as the product of probability distributions, designed for the individual parameter components.
For the components of the model parameter vector $\boldsymbol\theta$, which are the scaling factors for the parameter classes introduced in Sect. \ref{sec:swmmmodel}, we choose beta distributions with the mode equal to $1$ (or close to $1$ in the case of Manning coefficient for a pipe) and which vanish on the boundaries of the feasibility ranges specified in Sect. \ref{sec:swmmmodel}. The priors are centered at one, as we expect the authors of the model to have selected parameter values, which at least roughly correspond to reality. E.g. for the parameter \textit{slope of a subcatchment}, it is unlikely that it would be set to 50\% or 150\%  of the initial value, as it leads to obviously unrealistic response of the model.

For the variance of the measurement error, $\sigma^2_E$, we use a normal prior, based on previous measurements from the same site. For $\sigma^2_B$, we use an exponential prior distribution with a sharp decay, which expresses our desire that the data is explained foremost by the model and not by the bias correction term.
As the correlation factor $\tau$ is difficult to infer from data, we fix it with a delta function prior. Following \cite{giudice2013improving} we choose $1/3$ of the recession time, which, from Figure \ref{fig:design_}, is approximately $\tau=100$ min.



\subsection{Emulator of SWMM}


\paragraph{Simplified linear model}

As we have shown in \cite{emu_appli}, using a single linear reservoir as a prior for the emulator can already lead to a drastic improvement of its accuracy, compared to non-mechanistic emulators.
Since, typically, the sewer part of SWMM requires less calibration than the surface-runoff part, it seems reasonable to use surface parameters (1-7 in Sect. \ref{sec:swmmmodel}) alone to model this reservoir's retention time.
Using the same heuristic simplification of the Manning equation as in \cite{emu_appli}, we arrive at the following parametrization of the linear model (\ref{eqn:simpMod})

\begin{align}
    \frac {\mathrm d d(t)} {\mathrm d t} = -k\frac{w\sqrt{s}}{Anr}d(t)+p(t-t_0),
	\label{eqn:simplified_dynamics}
\end{align}
where $d(t)$ [m] is the state variable of the simplified system, namely the water level on the catchment surface, and $p(t)$ [m$\cdot$s$^ {-1}$] is the rainfall intensity.
The width of the overland flow path $w$ [m], the slope of the catchment $s$ [-], Manning's roughness coefficient of the catchment $n$ [m$^{-\frac 1 3}\cdot$s], and its imperviousness $r$ [\%] are averages derived from the corresponding SWMM parameters $\boldsymbol\theta$, weighted with the subcatchment's areas.

The parameters $k$ [m$^{\frac 2 3}$], $t_0$ [s] and $A$ [m$^2$] are components of the auxiliary parameter vector $\boldsymbol\theta'$ and need to be estimated.
The parameter $k$  is a linearization constant and $t_0$ is the lag of the catchment. Although we know the total area of the catchment, we prefer to rather use an estimate $A$ in the simplified model in order to partly make up for the SWMM components that are omitted from the simplified model (e.g.\ infiltration, evapotranspiration and the sewer part).

The discretized output of the simplified prior model is a flow time-series $y_i=Q_i$ [m$^3\cdot$s$^{-1}$], which is derived from \eqref{eqn:simplified_dynamics} and \eqref{eqn:projection_}, with $h(\boldsymbol\theta,\boldsymbol\theta')=Ar$, and reads

\begin{align}
        Q_i=k\frac{w\sqrt{s}}{n}d_i.
        \label{flowprojection}
\end{align}

As we emphasized in \cite{Machac}, the particular choice of parametrization in (\ref{eqn:simplified_dynamics}) is rather ad-hoc. 
We could also use the auxiliary parameter $k$ alone as a release rate and estimate it using the design data. But it is advantageous to employ some knowledge about parameter dependence with our heuristic release rate, which doesn’t express much more than an increase of the release rate if, on average, slopes are steeper, overland flow paths are wider, etc.
Data-driven methods of establishing optimal mappings between simulator and emulator parameters have been explored in \cite{Carbajal_2017}.

\paragraph{Design data generation}

We generate the design data so that they overreach the calibration bounds specified in Section \ref{sec:swmmmodel} $1.05\times$. This ensures sufficient accuracy at the boundary of the calibration space regardless of the prior probability.


\paragraph{Estimation of auxiliary emulator parameters}
From the auxiliary  parameters $\boldsymbol \theta' = (k,t_0,A,\gamma,\sigma)$, 
the first three are determined as described in Sect. \ref{sec:emulator} with their resulting values in Table \ref{tab:auxpar}. The choice of $\gamma$ proves to be not critical as values between 2 and 10 yield similarly good results. For this study, we have set $\gamma=5$, independently of the number of design data points. The noise parameter $\sigma$ is calculated according to eq. (\ref{eqn:sigmasq_estimate}), separately for 32, 64 and 128 design data points, for each of the three applications. From these values we derive the estimated RMSEs in Table \ref{tab:auxparsigma} as described in Sect. \ref{sec:results}. 

\pagebreak
\begin{table}
	\centering
\begin{tabular}{ c | c | c | c }
\hline
parameter & value (2) & value (4) &  value (8)\\ \hline\hline
$k$ [$\text{m}^{\frac{2}{3}}$ $\cdot10^{-7}]$ & 8.3 & 8.1 & 8.4 \\
$t_0$ [s] & 4 & 13 & 0\\
$A$ [m$^2\cdot10^6$] & 4.5 & 5.1  & 4.9 \\ 
 \hline
    \end{tabular}
    \caption{Estimated values of the auxiliary parameters for the 2, 4 and 8 parameter applications.}
    \label{tab:auxpar}
\end{table}

\begin{table}
	\centering
\begin{tabular}{ c | c | c | c }
\hline
\# d.d. & value (2) & value (4) &  value (8)\\ \hline\hline
32  & 9.54 (1.45) & 11.24 (1.73) & 20.11 (9.21) \\
64 & 7.79 (0.91) & 8.92 (1.02) & 15.02 (6.60)\\
128 & 5.83 (0.85) & 5.81 (0.88)  & 12.53 (3.11) \\ \hline
    \end{tabular}
    \caption{Estimated and measured (in parentheses) values of the RMSE, for the 2, 4 and 8 parameter applications, and for different sizes of the design data set.}
    \label{tab:auxparsigma}
\end{table}

\subsection{Results}\label{sec:results}

\begin{figure}
	\centering
	\hspace*{-1cm}
	\includegraphics[scale=0.45]{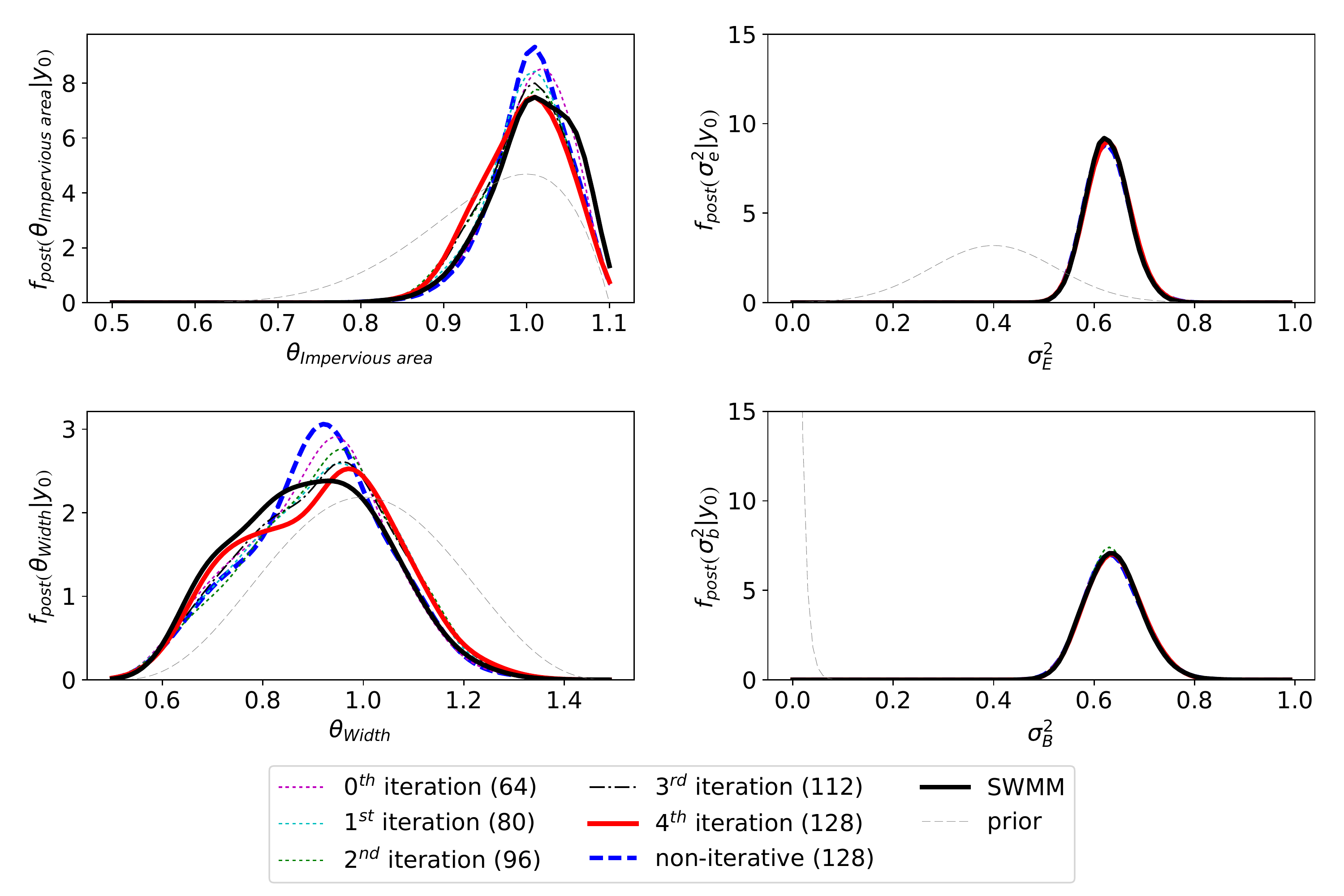}
	\caption{Comparison of the posterior marginals for 2 SWMM parameters (+2 error model parameters), obtained with SWMM and with the emulator, with and without the iterative improvement. We have used 128 design data points in total. In the case of the iteratively improved emulator, we have sampled 64 design data points with the Halton sequence and then added 16 at each iteration step.  }
	\label{fig:pos2}
\end{figure}

\begin{figure}
	\centering
	\includegraphics[scale=0.45]{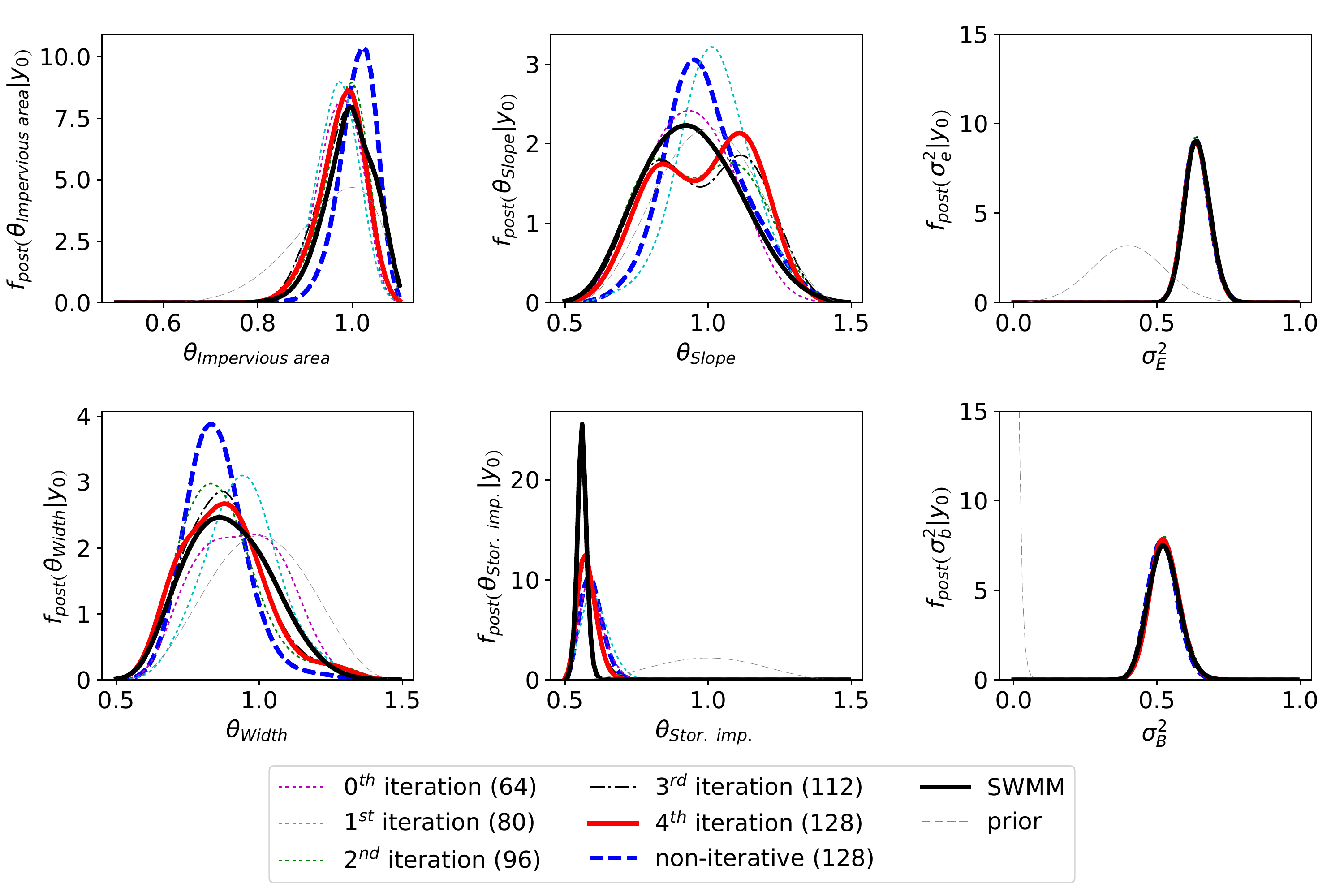}
    \caption{Analogous to Figure \ref{fig:pos2}, but for the 4 parameter application. }
	\label{fig:pos4}
\end{figure}

\begin{figure}
	\centering
    \vspace*{-4cm}
	\includegraphics[scale=0.45]{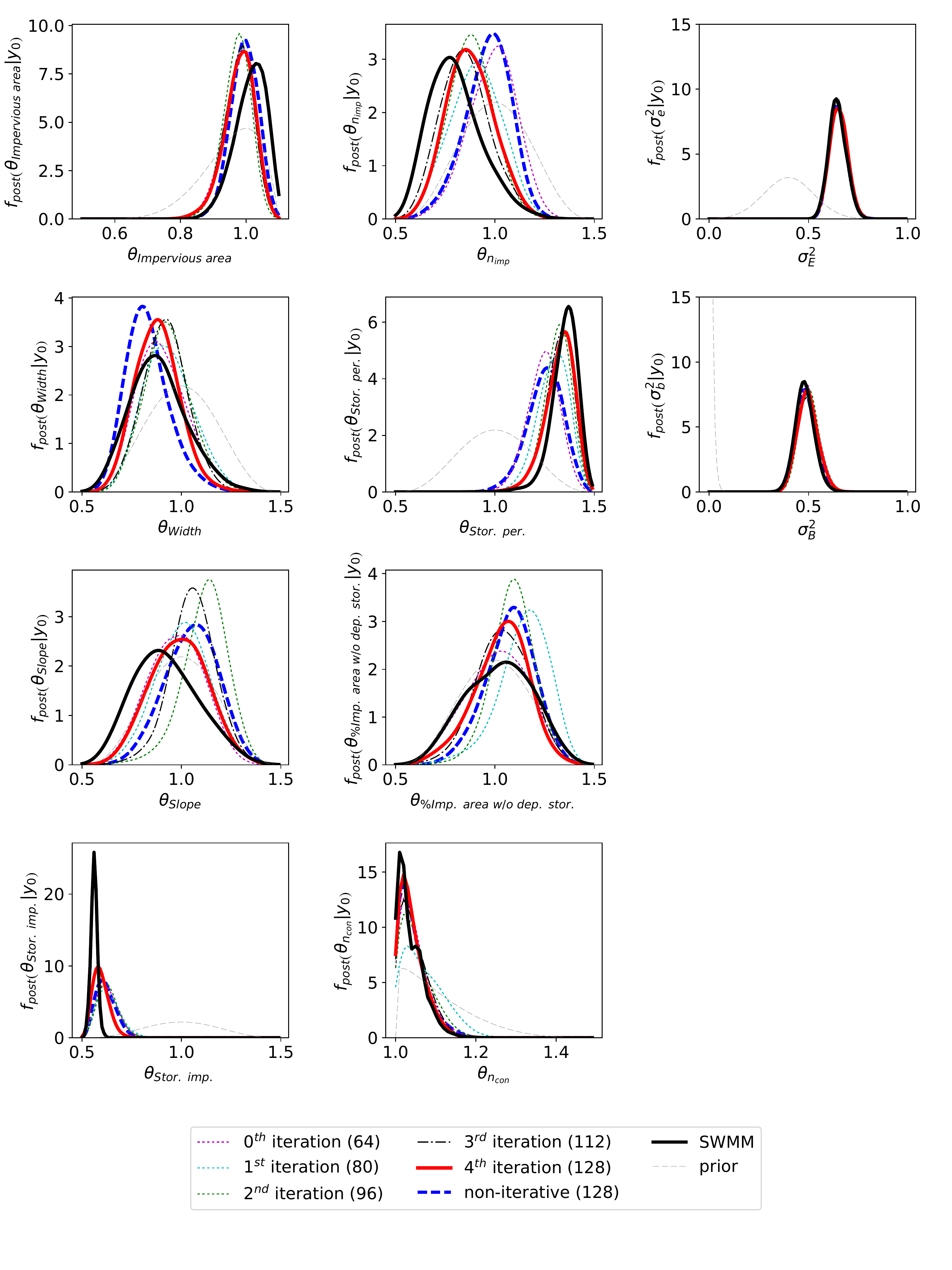}
    \caption{Analogous to Figure \ref{fig:pos2}, but for the 8 parameter application. }
	\label{fig:pos8}
\end{figure}

In Figs.\ \ref{fig:pos2}, \ref{fig:pos4} and \ref{fig:pos8} we show the marginals of the posterior distributions for inference of 2, 4 and 8 SWMM parameters jointly with 2 parameters of the error model, acquired with SWMM and with different strategies of choosing design data points for the emulator.
In line with previous findings \citep{emu_appli}, the results show that, unless an iterative scheme is used, 128 design data points do not always lead to a significantly better result than 64 points. When only 2 or 4 parameters are inferred, the results are already good with only 64 design data points (Figs.\ \ref{fig:pos2} and \ref{fig:pos4}). Doubling the number of points does not improve the results unless an iterative scheme is employed. Whereas in the 4 parameter application the iterative scheme improves the results only slightly, the improvement is bigger in the 8 parameter application (Fig.\ \ref{fig:pos8}).
Despite the more significant improvement in the 8 parameter case, already the marginals indicate that we do not get an accurate representation of the posterior.
This confirms the guess that we can hardly expect a very high emulation accuracy in an 8 dimensional parameter space with just 128 design data points.
In fact, the results are remarkably good given this small size of the design data set that clearly reaches its limits for an 8 dimensional parameter space.

\begin{figure}
	\centering
	\includegraphics[scale=0.40]{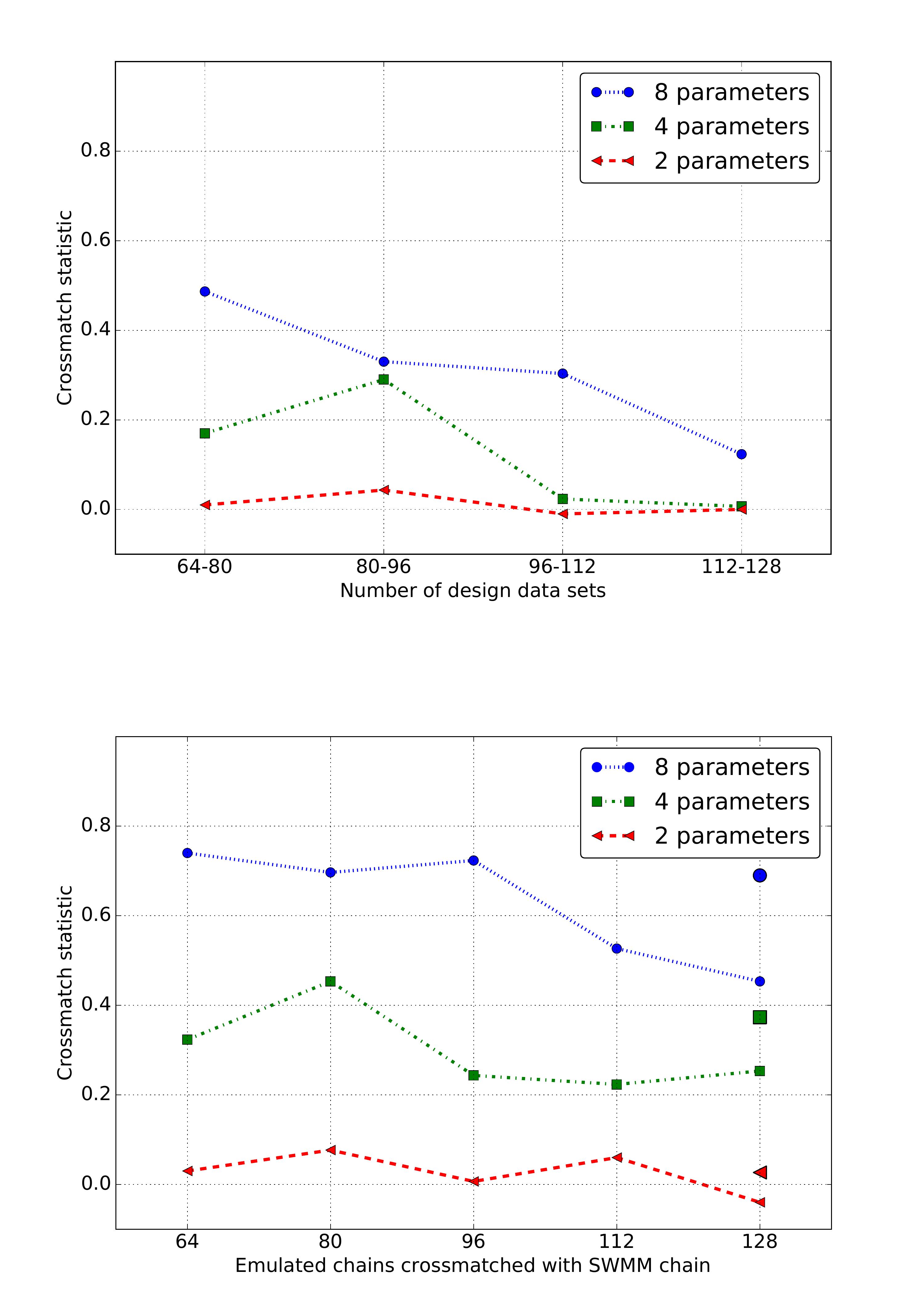}
    \caption{Top panel: Decrease in difference \eqref{equ:crossmatchdistance} between successive iterative distributions with increasing number of iterations (labels indicate design set sizes of the two distributions that are compared). Bottom panel: (Slow) approximation of the iterative distributions to the posterior distribution produced with the full model (labels indicate design set sizes). The isolated point at a design set size of 128 represents results for a non-iterative design and illustrates the superiority of the iterative approach.}
	\label{fig:crossmatch}
\end{figure}

Figure \ref{fig:crossmatch} shows the results for the distance measure \eqref{equ:crossmatchdistance} for subsequent, iterative samples (top panel) and for iterative samples compared to the sample from the posterior without using the emulator (bottom panel).
The red line in the top panel clearly indicates that iteration does not considerably modify the sample, for the 2 parameter application. The red line in the bottom panel reveals the cause of this observation: already the first sample is very close to the posterior without using the emulator. There is no need for iteration in the 2 parameter application.
The green line in the top panel of Figure \ref{fig:crossmatch} shows a decreasing trend in the degree the iterative distributions are modified in the 4 parameter application and the green line in the bottom panel shows that the iterative distributions get closer to the one calculated without emulator for this case.
The final distribution of the iterative process is still not identical to the posterior calculated without emulator, but it is much closer to it than the one using the same design data set size without iterative refinement.
Finally, the blue lines show the same trend as the green lines for the 8 parameter application although considerably more pronounced as in the 4 parameter application, but also with a larger final distance from the posterior calculated without emulation.
Also here, the final distribution of the iterative process is considerably closer to the posterior calculated without using the emulator than the posterior calculated with an emulator based on the same number of design data points in a non-iterative setting.

\begin{figure}
	\centering
	\vspace*{-2.5cm}
    \includegraphics[scale=0.45]{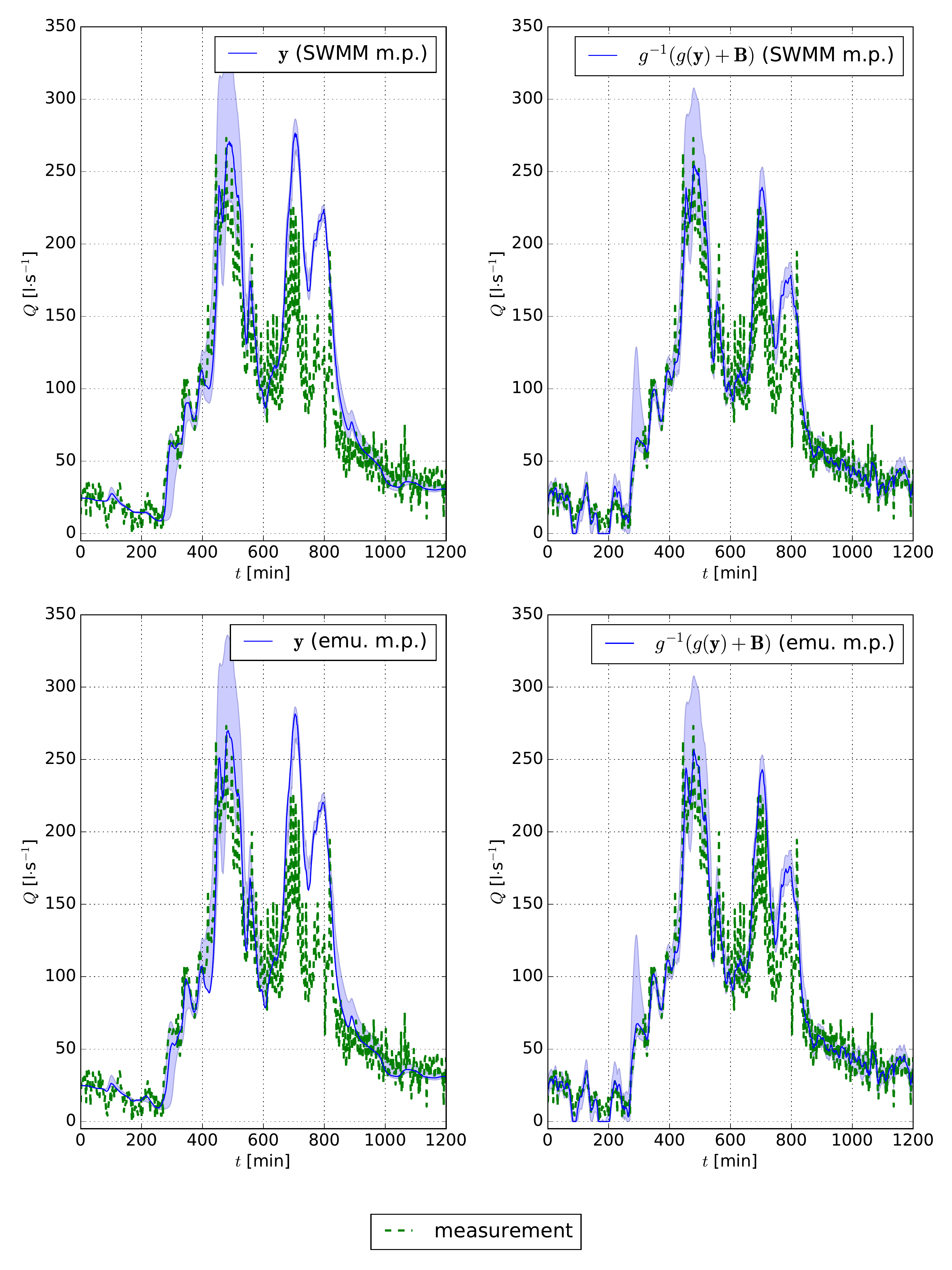}
    \caption{This figure shows, on the left side, outputs of a SWMM run for the maximum posterior obtained with SWMM and of an emulator run for the maximum posterior obtained with the emulator. The plots on the right side are analogous, but before we apply the inverse Box-Cox transformation, we add the mean of the bias (see \cite{Reichert2012} for details). The $95\%$ predictive uncertainty bands are shown, for the parametric uncertainty (left) and parametric uncertainty plus bias (right).}
	\label{fig:result}
\end{figure}

Whilst the posterior on this 8-dimensional parameter space obtained with the emulator differs from the one obtained with SWMM, Fig. \ref{fig:result} shows that the effect of this difference on the output distributions in the calibration phase is not noticeable. The left panels show the effect of the parametric uncertainty on the output of the deterministic model (without bias correction) and the right panels show the output of the stochastic model (with bias correction) at the maximum of the posterior. Fig. \ref{fig:result} also demonstrates the importance of using an adequate likelihood function in cases with significant bias, which are typical for hydrological applications.

In Table \ref{tab:times}, we show a comparison of conditioning times, emulation times and the times needed to evaluate $(\sigma_E^2\boldsymbol 1+\Sigma_B)^{-1}(g(\mathbf y_o)-g({\mathbf y}(\boldsymbol\theta)))$, for various amounts of design data. The last item is required for the evaluation of the log-likelihood function \eqref{eqn:lhood}, and the times are to be understood without the model run.
Here, we have used a Kalman filter, for the evaluation of the emulator \citep{Reichert2011}, whose computational complexity scales cubically with the number of design data points.
For large design data sets, we recommend using the algorithm presented in \cite{Albert}, which scales linearly with the number of design data points, at the cost of a computationally more costly conditioning phase. This conditioning phase can become a limiting factor, if both the number of design data points and the output dimension become large. In these cases further numerical improvements such as covariance matrices with compact support \citep{kaufman2011} will be required.

Finally, in Table \ref{tab:auxparsigma}, we compare estimated and measured RMSEs of the emulator. The former are derived from the diagonal elements of the covariance matrix (\ref{conditionedcov}). Both the estimated and the measured RMSEs increase with increasing number of parameters and decrease with increasing number of design data points. The results also show that the actual errors are even considerably smaller than the predicted ones.

\begin{table}
	\centering
\begin{tabular}{ c | c | c | c }
\hline
\# d.d. & cond. $t$ [ms] & emu. $t$ [ms] &  $(\sigma_E^2\boldsymbol 1+\Sigma_B)^{-1}(g(\mathbf y_o)-g({\mathbf y}(\boldsymbol\theta)))$ $t$ [ms]\\ \hline\hline
32 & 130 & 45 & \\
64 & 794 & 229 & 90\\
128 & 5780 & 1480 & \\
 \hline
    \end{tabular}
    \caption{Overview of the computation times needed for conditioning and emulation, for various amounts of design data points. The evaluation time of the log-likelihood function (the last column) does not depend on the number of design data points. However, for small numbers, it constitutes a significant share of the total time needed for Bayesian inference. In comparison, a SWMM run takes 19 seconds on average.}
    \label{tab:times}
\end{table}

\clearpage

\section{Summary and conclusions}

Bayesian parameter inference typically requires a large number of model runs, particularly if the dimension of the parameter space is high. If a model run is computationally expensive, which is often the case in the environmental sciences, this poses a formidable task.

We propose a Bayesian inference technique that keeps the number of required model runs low by combining three strategies: (i) the use of a mechanistic emulator, (ii) low-discrepancy sampling of the parameter space, and (iii) iterative refinement of the design data set in regions of the parameter space that are relevant for the posterior.
For a case study from urban hydrology, we demonstrated that we get good accuracy of the posterior for 2 or 4 emulated parameters and still reasonable accuracy for 8 parameters with only 128 simulator runs.

While this may be an extreme example, we can still expect that the suggested technique can save considerable simulation time even if more simulations will be used.
Due to the unfavorable scaling of the employed emulation technique with the number of design data points, this may need further numerical improvements or the use of a different emulator.

\section*{Appendix}
\appendix

\section{Mathematical notation}
          Where applicable, we show the number of an equation, which helps to understand the meaning of the symbol the most. Generally, bold symbols mean either a vector or a matrix. We do not list the bold versions, unless the object appears only as a vector or as a matrix.

\begin{table}[H]
\begin{tabular}{ l  l }
$A$& Area of a catchment\\
$\mathbf B$& Additive bias correction term \eqref{model}\\
$\mathbf C$& Coupling matrix \eqref{eqn:noiseterm}\\
$d$& General state variable \eqref{eqn:simpMod} or a water level on catchment's surface \eqref{eqn:simplified_dynamics}\\
$d_{cm}$& Cross-match distance \eqref{equ:crossmatchdistance}\\
$\mathbf E$& Measurement error \eqref{model}\\
$f_{\text{post}}$& Posterior probability density function
	\eqref{eqn:bayes}\\
$f_{\text{prior}}$&Prior probability density function \eqref{eqn:bayes}\\
$g$& Box-Cox transformation function \eqref{model}\\
$G$& Green's function \eqref{eqn:Sigmacoupled}\\
$h$& Output function \eqref{eqn:projection_}\\
$k$& Linearization constant \eqref{eqn:simplified_dynamics}\\
$l$& Likelihood function \eqref{eqn:lhood}\\
$n$& Integer denoting either a sample size or the number of parameters in various contexts or\\
&Manning's roughness coefficient \eqref{eqn:simplified_dynamics}\\
$n_{cm}$& Number of cross-matches \eqref{equ:crossmatchdistance}\\
$N_t$& Length of a measured time series\\
$p$& General linear system input function \eqref{eqn:simpMod} or Rainfall intensity  \eqref{eqn:simplified_dynamics}\\
$r$& Imperviousness \eqref{eqn:simplified_dynamics}\\
$\mathbf R$& Correlation function \eqref{eqn:cor_3}\\
$\mathbb R^{n+1}$& $n+1$-th dimensional space of real numbers\\
$Q$& Flow \eqref{flowprojection}\\
 \end{tabular}
 \end{table}
 \begin{table}[!ht]
\begin{tabular}{ l  l }
$s$& Slope of a catchment \eqref{eqn:simplified_dynamics}\\
$t_i$& Discrete time points\\
$t_0$& Lag of a catchment \eqref{eqn:simplified_dynamics}\\
$w$& Width of overland flow  \eqref{eqn:simplified_dynamics}\\
$\mathbf y$& Deterministic model output \\
$\bar {\mathbf y}$&Mean emulator output \eqref{conditionedmean}\\
$\mathbf y_o$& Vector of measured data\\
$\mathbf Y$& Output of a dynamical model in the form of a random vector \eqref{model} \\
$z$& Mean of a coupled prior linear model (before conditioning) \eqref{eqn:Sigmacoupled}\\
$\gamma$& Correlation length \eqref{eqn:cor_3}\\
$\delta^{\alpha}_{\beta}$& Kronecker's delta\\
$\boldsymbol{\eta}$ & Gaussian white noise \eqref{eqn:coupled_sys}\\
$\theta_{p1}$& Parameter \textit{p1}\\
$\boldsymbol\theta$& Vector of model parameters\\
$\boldsymbol\theta'$& Auxiliary emulator parameters \eqref{eqn:simpMod}\\
$\boldsymbol\theta^*$& Stretched parameter vector \eqref{eqn:stretch}\\
$\widetilde{\boldsymbol{\theta}}$&Parameter vectors associated with the $n+1$ replica \eqref{eqn:coupled_sys}\\
$\kappa$&Linear system coefficient function \eqref{eqn:simpMod}\\
$\lambda$& Box-Cox transformation parameter or Stretch parameter \eqref{eqn:stretch}\\
$\mu$& Parameter mean of design data \eqref{eqn:stretch}\\
${\rho_l}$& Normalizing constant \eqref{eqn:cor_3}\\
$\sigma$& Emulator noise standard deviation \eqref{eqn:noiseterm}\\
$\sigma_E$& Measurement error term standard deviation \eqref{model}\\
$\sigma_B$& Bias term standard deviation \eqref{eqn:bias}\\
$\Sigma_{B,i,j}$& Bias covariance matrix \eqref{eqn:bias}\\
$\Sigma$& Covariance matrix of coupled prior linear model \eqref{eqn:Sigmacoupled}\\
$\Sigma'$& Covariance matrix associated only with the first $n$ replica \eqref{conditionedcov}\\
$\overline\Sigma$& Covariance matrix the emulator \eqref{conditionedcov}\\
$\tau$& Auto-correlation time  \eqref{eqn:bias}\\
$\boldsymbol 1$& Unit matrix
 \end{tabular}
 \end{table}
\newpage
 \section*{Data and Software}

Software used to generate the data used in this article, as well as the data, is available at \url{https://github.com/machacd/mechemu}. The data can also be downloaded separately from \url{https://www.dropbox.com/s/fomqodwd9vwp1ec/resulting_samples.zip?dl=0}.

\section*{Acknowledgements}

This work is funded by the Swiss National Science Foundation, grants no. CR22I2 135551 and CR22I2 152824 in scope of the project ``Using Commercial Microwave Links and Computer Model Emulation to Reduce Uncertainties in Urban Drainage Simulations'' (COMCORDE). The authors would also like to thank Tobias Doppler for providing us with all the measurements used in this work and to Juan Pablo Carbajal for his valuable comments.

\bibliographystyle{abbrvnat}
\bibliography{library}

\end{document}